\documentclass[aps,pra,twocolumn,superscriptaddress,nofootinbib]{revtex4-1}
\usepackage{amsmath}
\usepackage{amssymb}
\usepackage{graphicx}
\usepackage{mathrsfs}
\usepackage{ntheorem}
\usepackage{mathtools}
\usepackage{enumerate}
\usepackage{enumitem}
\usepackage{times,txfonts}
\usepackage{subfigure}
\usepackage{upgreek}
\usepackage{tikz}
\usepackage[colorlinks]{hyperref}
\newcommand{\ket}[1]{|#1\rangle}
\newcommand{\bra}[1]{\langle #1|}
\newcommand{\Tr}{\mathrm{Tr}}

\newcommand{\abs}[1]{\lvert #1\rvert}

\def\CC{{\rm\kern.24em \vrule width.04em height1.46ex depth-.07ex \kern-.30em C}}
\def\RR{{\rm\kern.24em \vrule width.04em height1.46ex depth-.07ex
\kern-.30em R}}
\def\P{{\rm I\kern-.25em P}}

\begin{document}

\title{Increasing the dimension of the maximal pure coherent subspace of a state via incoherent operations}

\author{C. L. Liu}

\affiliation{Institute of Physics, Beijing National Laboratory for
  Condensed Matter Physics, Chinese Academy of Sciences, Beijing
  100190, China}
\affiliation{Graduate School of China Academy of Engineering Physics, Beijing 100193, China}
\author{D. L. Zhou}

\email{zhoudl72@iphy.ac.cn}

\affiliation{Institute of Physics, Beijing National Laboratory for
  Condensed Matter Physics, Chinese Academy of Sciences, Beijing
  100190, China}

\affiliation{School of Physical Sciences, University of Chinese
  Academy of Sciences, Beijing 100049, China}

\affiliation{CAS Central of Excellence in Topological Quantum
  Computation, Beijing 100190, China}

\affiliation{Songshan Lake Materials Laboratory, Dongguan, Guangdong
  523808, China}

\date{\today}

\begin{abstract}
Quantum states transformation under free operations plays a central
role in the resource theory of coherence. In this paper, we investigate
the transformation from a mixed coherent state into a pure one by
using both incoherent operations and stochastic incoherent
operations. We show that contrary to the strictly incoherent operations
and the stochastic strictly incoherent operations, both the incoherent
operations and the stochastic incoherent operations can
increase the dimension of the maximal pure coherent subspace of a state.
This means that the
incoherent operations are generally stronger than the strictly incoherent
operations when we want to transform a mixed coherent state into a pure coherent
one. Our findings can also be interpreted as confirming the ability of
incoherent operations to enhance the coherence of mixed states relative
to certain coherence monotones under strictly incoherent operations.
\end{abstract}

\maketitle

\section{Introduction}

Quantum coherence is an important feature of quantum mechanics that
is responsible for the departure between the classical and the quantum
world. It is an essential component in quantum information processing
\cite{Nielsen}, and it plays a central role in emergent fields, such as
quantum metrology \cite{Giovannetti,Giovannetti1}, nanoscale
thermodynamics \cite{Aberg,Lostaglio,Lostaglio1}, and quantum biology
\cite{Sarovar,Lloyd,Huelga,Lambert}. Recently, the quantification of
coherence has attracted a growing interest due to the rapid
development of quantum information science
\cite{Aberg1,Baumgratz,Levi,Streltsov,Fan}.

As a quantum resource
theory, there are two fundamental ingredients: free states and free
operations \cite{Chitambar,Brandao,Liu3}. For the resource theory of
coherence, the free states are quantum states which are diagonal in a
prefixed reference basis. While, there are no general consensus on the
set of free operations \cite{Streltsov,Fan}. Based on various physical
and mathematical considerations, several free operations were
presented, such as the maximally incoherent operations (MIO) \cite{Aberg1},
the incoherent operations (IOs) \cite{Baumgratz}, the strictly incoherent
operations (SIOs) \cite{Winter,Yadin}, and the physically incoherent
operations (PIOs) \cite{Chitambar2,Chitambar1}. Here, we focus
our attention on the IOs, which have the physical
motivation that they cannot create coherence \cite{Baumgratz}, and the SIOs,
 which have the physical motivation that they can neither
create nor use coherence \cite{Yadin}.

With these notions, a central topic of the resource theory of
coherence is to study the states transformation under free operations.
Investigations on this topic started from the deterministic
transformation between pure coherent states. In Refs. \cite{Du,Zhu},
the authors presented the necessary and sufficient conditions for the
deterministic transformation between pure coherent states by using
incoherent operations. Then, in Ref. \cite{Chitambar1}, the author
present that any pure state transformation by using strictly
incoherent operations obeys the same necessary and sufficient
conditions for the deterministic transformation of an incoherent
operations. These results tell us that when we want to transform a
pure state into another one deterministically, incoherent operations
and strictly incoherent operations have the same power. After that,
the probabilistic transformation between pure coherent states was
studied in Refs. \cite{Du1,Zhu}. Their results also show that the
incoherent operations and strictly incoherent operations have the same
power. More recently, we know that the power of incoherent operations
and strictly incoherent operations is identical when performing the
deterministic transformation between $2$-dimensional mixed states
\cite{Chitambar1,Shi,Streltsov1}. Despite these results, one might still
suspect that incoherent operations are more powerful than strictly
incoherent operations in general.

In this paper, we show that the above conjecture is correct, i.e., we find that the
incoherent operations are stronger than strictly incoherent operations when
transform a mixed coherent state into a pure coherent state.
Specifically, we will show that contrary to the strictly incoherent operations
and the stochastic strictly incoherent operations, both the incoherent operations
and the stochastic incoherent operations can  increase the dimension of the
maximal pure-coherent subspace of a state. Our findings imply that there is
indeed an operational gap
between incoherent operations and strictly incoherent operations
under state transformations, which provide an answer to the open question in
Ref. \cite{Chitambar1}. An interesting consequence of this results further imply there exist
coherence monotones under strictly incoherent operations that can increase
under incoherent operations.

This paper is organized as follows. In Sec.~II, we recall some notions
of the quantum resource theory of coherence, including incoherent
operations, strictly incoherent operations, stochastic incoherent
operations, and stochastic strictly incoherent operations. In
Sec.~III, we show our main results, i.e., show that the incoherent
operations are generally stronger than strictly incoherent operations
when we want to transform a mixed coherent state
into a pure one. A concise summary of our results is presented in Sec.~IV.

\section{Preliminaries}

Let $\mathcal {H}$ be the Hilbert space of a $d$-dimensional quantum
system. A particular basis of $\mathcal {H}$ is denoted as
$\{\ket{i}, ~i=1,2,\cdot\cdot\cdot,d\}$, which is chosen according to the physical
problem under consideration. Coherence of a state is then measured
based on the basis chosen. Specifically, a state is said to be
incoherent if it is diagonal in the basis. Any state which cannot be
written as a diagonal matrix is defined as a coherent state. For a
pure state $\ket{\varphi}$, we will denote
$\ket{\varphi}\bra{\varphi}$ as $\varphi$, i.e.,
$\varphi:=\ket{\varphi}\bra{\varphi}$ for the sake of simplicity.

An incoherent operation \cite{Baumgratz} is a completely positive
trace-preserving map, expressed as
\begin{eqnarray}
  \Lambda(\rho)=\sum_{n=1}^N K_n\rho K_n^\dagger,
\end{eqnarray}
where the Kraus operators $K_n$ satisfy not only
$\sum_n K_n^\dagger K_n= \mathbb{I}$ but also
$K_n\mathcal{I}K_n^\dagger\subset \mathcal{I}$ for all $K_n$, i.e., each
$K_n$ transforms an incoherent state into an incoherent state and such
a $K_n$ is called an incoherent Kraus operator. While, a strictly
incoherent operation \cite{Winter, Yadin} is a completely positive
trace-preserving map satisfying not only
$\sum_n K_n^\dagger K_n= \mathbb{I}$ but also
$K_n\mathcal{I}K_n^\dagger\subset \mathcal{I}$ and
$K_n^\dagger\mathcal{I}K_n\subset\mathcal{I}$ for all $K_n$, i.e., each
$K_n$ as well $K_n^\dagger$ transforms an incoherent state into an
incoherent state and such a $K_n$ is called a strictly incoherent
Kraus operator. Here, $\mathcal{I}$ represents the set of incoherent
states.

With the notions of the incoherent operation and the strictly
incoherent operation, we can introduce the notion of the stochastic
incoherent operation \cite{Bu} and the stochastic strictly incoherent
operation \cite{Liu}. A stochastic incoherent operation is constructed
by a subset of incoherent Kraus operators. Without loss of generality,
we denote the subset as $\{K_{1},K_{2},\dots, K_{L}\}$. Otherwise, we
may renumber the subscripts of these Kraus operators. Then, a
stochastic incoherent operation, denoted as $\Lambda_s(\rho)$, is defined by
\begin{equation}
  \Lambda_s(\rho)=\frac{\sum_{n=1}^L K_{n}\rho K_{n}^{\dagger}}{\Tr(\sum_{n=1}^LK_{n}\rho K_{n}^{\dagger})},
  \label{lams}
\end{equation}
where $\{K_{1},K_{2},\dots, K_{L}\}$ satisfies
$\sum_{n=1}^L K_{n}^{\dagger}K_{n}\leq I$. Clearly, the state
$\Lambda_s(\rho)$ is obtained with probability
$P=\Tr(\sum_{n=1}^LK_{n}\rho K_{n}^{\dagger})$ under a stochastic incoherent
operation $\Lambda_s$, while state $\Lambda(\rho)$ is fully deterministic under an
incoherent operation $\Lambda$. Similarily, we can give the notion of the
stochastic strictly incoherent operation by changing the incoherent
Kraus operator into the strictly incoherent Kraus operator in Eq.
(\ref{lams}).

A functional $C$ can be taken as a measure of coherence if it
satisfies the four postulate \cite{Baumgratz,Yadin}: (C1) the
coherence being zero for incoherent states; (C2) the monotonicity of coherence under incoherent
operations or strictly incoherent operations; (C3) the monotonicity of
coherence under selective measurements on average; and (C4) the
non-increasing of coherence under mixing of quantum states. A coherence monotone satisfies (C1)
and (C2), while a coherence measure satisfies (C1)-(C4).   In
accordance with the general criterion, several coherence measures have
been put forward. Out of them, we recall the coherence rank
\cite{Winter}, which will be considered in this paper. The coherence rank
$C_r$ of a pure state (not necessarily normalized),
$\ket{\varphi}=\sum_{i=1}^Rc_i\ket{i}$ with $c_i\neq 0$, is defined as the number
of nonzero terms of $c_i\neq 0$ minus $1$, i.e.,
\begin{eqnarray}
  C_r(\varphi)=R-1.
\end{eqnarray}
For a mixed state $\rho$, the coherence rank of it is defined as
$C_r(\rho)=\inf_{\{p_i,\varphi_i\}}\sum_ip_iC_r(\varphi_i)$, where
$\rho=\sum_ip_i\varphi_i$ is any decomposition of $\rho$ into pure states
$\varphi_i$ with $p_i\geq0$.

\section{Incoherent operations being stronger than strictly incoherent operations}

We begin our study by observing the difference between the structure of incoherent
operations and that of strictly incoherent operations from their Kraus
operators. This leads to the following Lemma \cite{Du,Winter,Yao}.

\emph{Lemma} 1.-- (a) For an incoherent Kraus operator, there is at
most one nonzero element in each column of $K_n$;\\
(b) For a strictly incoherent Kraus operator, there is at most one
nonzero element in each column and each row of $K_n$.

Equipped with the above Lemma, we now present the following theorem.

\emph{Theorem} 1.-- Let $\rho$ be any $2$-dimensional mixed states. Then
for any coherent state $\varphi$, no stochastic incoherent operation can
transform $\rho$ into $\varphi$ with nonzero probability.

\emph{Proof}.--First, we show that if we want to judge whether there
exists a stochastic incoherent operation such that
\begin{eqnarray}
  \Lambda_s(\rho)=\varphi,
\end{eqnarray}
we only need to consider the stochastic incoherent operation with the
form of
\begin{eqnarray}
  \Lambda_s^1(\rho)=\frac{K\rho K^\dag}{\Tr(K\rho K^\dag)}. \label{Step1}
\end{eqnarray}

To this end, we assume that we can transform a mixed state $\rho$ into a
pure coherent state $\varphi$ by using some stochastic incoherent operation
$\Lambda_s$, i.e.,
\begin{equation}\nonumber
  \Lambda_s(\rho)=\frac{\sum_{n=1}^L K_{n}\rho K_{n}^{\dagger}}{\Tr(\sum_{n=1}^LK_{n}\rho K_{n}^{\dagger})}=\varphi.\end{equation}
Then, since pure states are extreme points of the set of states, there must be
\begin{eqnarray}
  \frac{K_{n}\rho K_{n}^{\dagger}}{\Tr(K_{n}\rho K_{n}^{\dagger})}=\varphi
\end{eqnarray}
for all $n=1,...,L$. On the other hand, we note that
\begin{equation}\nonumber
  \Lambda_s^1(\rho)=\frac{K_n\rho K_n^\dag}{\Tr(K_n\rho K_n^\dag)}
\end{equation}
is also a stochastic incoherent operation. Thus, we obtain that if we
can transform $\rho$ into $\varphi$ by using $\Lambda_s^1(\rho)$, then there exists a
stochastic incoherent operation such that $\Lambda_s(\rho)=\varphi$.

With the above result, to prove the theorem, it is enough to examine
all the stochastic incoherent operations with the forms
\begin{eqnarray}
  \Lambda_s^1(\rho)=\frac{K\rho K^\dag}{\Tr(K\rho K^\dag)},
\end{eqnarray}
where $K$ being a $2\times2$ incoherent Kraus operators.

From the definition of incoherent operations, we can get that there
are eight classes of incoherent Kraus operators with the form
\begin{eqnarray}
  K_1&&=\begin{pmatrix}
    a_1&0\\
    0&0
  \end{pmatrix},~~ K_2=\begin{pmatrix}
    0&a_2\\
    0&0
  \end{pmatrix},\nonumber\\
  K_3&&=\begin{pmatrix}
    0&0\\
    a_3&0
  \end{pmatrix},~~ K_4=\begin{pmatrix}
    0&0\\
    0&a_4
  \end{pmatrix}, \nonumber\\
  K_5&&=\begin{pmatrix}
    a_5&b_5\\
    0&0
  \end{pmatrix},~~ K_6=\begin{pmatrix}
    a_6&0\\
    0&b_6
  \end{pmatrix},\nonumber\\
  K_7&&=\begin{pmatrix}
    0&0\\
    a_7&b_7
  \end{pmatrix},~~ K_8=\begin{pmatrix}
    0&b_8\\
    a_8&0
  \end{pmatrix},
\end{eqnarray}
where $a_n\neq0$ and $b_n\neq0$ for all $n=1,\cdots,8$. It is straightforward to
see that $K_1$, $K_2$, $K_3$, $K_4$, $K_5$, and $K_7$ cannot transform
any mixed states into a pure coherent state with nonzero probability.
Thus, we only need to examine the $\Lambda_s^1(\rho)=K_n\rho K_n^\dag$ with
$n=6,8$. However, since $K_6$ and $K_8$ are strictly incoherent Kraus
operators, by using a result in Ref. \cite{Liu}, which says a
$d$-dimensional mixed state $\rho$ can never be transformed into a pure
coherent state with its coherence rank being $d-1$ by using a
stochastic strictly incoherent operation, we immediately obtain that
we cannot transform any $2$-dimensional mixed state $\rho$ into a pure
coherent state $\varphi$ by using stochastic incoherent
operations. This completes the proof of the theorem.~~~~~~~~~~~~~~~~~~~~~~~~~~~~~~~~~~~~~~~~~~~~~~~~~~~~~~
~~~~~~~~~~~~~~~~~~~~~~~~~~~~~~~~~~~~~~~~~~~~~~$\blacksquare$

With the above theorem, we immediately have the following corollary:

\emph{Corollary} 1.--Let $\rho$ be a $2$-dimensional mixed state. Then, for any coherent state
$\ket{\varphi}$, no incoherent operation can transform $\rho$ into $\varphi$ with probability one.

We now move onto the larger dimensional state space of $3$
and show that stochastic incoherent operations and stochastic strictly
incoherent operations are no longer the same. This arrive at the following theorem.

\emph{Theorem} 2.-- Let $\rho$ be any $3$-dimensional mixed state. If there
is no incoherent projector $\mathbb{P}$ such that
\begin{eqnarray}
  \frac{\mathbb{P}\rho\mathbb{P}}{\Tr(\mathbb{P}\rho\mathbb{P})}=\varphi,\label{Theorem2}
\end{eqnarray}
with the coherence rank of $\ket{\varphi}$ being equal to or greater than
$1$. Then, (a) there are $3$-dimensional mixed state $\rho$ and pure coherent
  state $\varphi$ such that the transformation from $\rho$ into
  $\varphi$ can be achieved by using stochastic incoherent operations;
(b) it is impossible to transform $\rho$ into a pure coherent
  state $\varphi$ by using stochastic strictly incoherent operations.

Here, an incoherent projector is a projector which has the form
$\mathbb{P}_{\text{I}}=\sum_{i\in{\text{I}}}\ket{i}\bra{i}$ with
$\text{I}\subset\{1,...,d\}$.

\emph{Proof.}--(a) Let us consider a special class of incoherent Kraus
operators with the form
\begin{eqnarray}
  \widetilde{K}=k~\begin{pmatrix}
    a_1&a_2&0\\
    0 & 0&a_3\\
    0& 0& 0
  \end{pmatrix},
\end{eqnarray}
where $a_1,a_2,a_3$ are all nonzero complex numbers and $k$ is some
complex number such that
$\widetilde{K}^\dag \widetilde{K}\leq\mathbb{I}$. We further choose
$\rho=p_1\varphi_1+p_2\varphi_2$ with
\begin{eqnarray}
  \ket{\varphi_1}&&=\varphi_1^1\ket{1}+\varphi_2^1\ket{2}+\varphi_3^1\ket{3}\nonumber\\
  \ket{\varphi_2}&&=\varphi_1^2\ket{1}+\varphi_2^2\ket{2}+\varphi_3^2\ket{3}.
\end{eqnarray}
Suppose that we can transform $\rho$ into a pure state $\ket{\varphi}$ by using
some stochastic incoherent operations if and only if there are
\begin{eqnarray}
  \widetilde{K}\ket{\varphi_1}=k_1\ket{\varphi},~~
  \widetilde{K}\ket{\varphi_2}=k_2\ket{\varphi},\label{decomposition}
\end{eqnarray}
where $k_1$ and $k_2$ may be identical or not identical since we only consider the
transformation in Eq. (\ref{Theorem2}).
Here we should note that this condition is independent of the ensemble
decomposition of $\rho$. To see this, assume that
$\rho=\sum_{i=1}^n\uplambda_i\ket{\uplambda_i}\bra{\uplambda_i}$ is another
ensemble decomposition of $\rho$. Then, we can transform $\rho$ into a pure
state $\varphi$ by using an incoherent operator $\widetilde{K}$ if and only
if there are
\begin{eqnarray}
  \widetilde{K}\ket{\uplambda_i}=k_i\ket{\varphi}, \label{condition_2}
\end{eqnarray}
for all $i=1,\cdots,n$ with $k_i$ being complex numbers. On the other hand,
$\{p_i,\ket{\varphi_i}\}$ is an ensemble for $\rho$ if and only if there exists
a unitary matrix $U=\left(U_{ij}\right)$ such that \cite{Hughston}
\begin{eqnarray}
  \sqrt{p_i}\ket{\varphi_i}=\sum_jU_{ij}\sqrt{\uplambda_j}\ket{\uplambda_j}.
\end{eqnarray}
By using the condition in Eq. (\ref{condition_2}), we then obtain that
\begin{eqnarray}
  \widetilde{K}\ket{\varphi_i}&&=\frac1{\sqrt{p_i}}\sum_jU_{ij}
  \sqrt{\uplambda_j}\widetilde{K}\ket{\uplambda_j}=\sum_j\frac{U_{ij}
  \sqrt{\uplambda_j}}{\sqrt{p_i}}k_j\ket{\varphi}\nonumber\\&&=
  k_i^\prime\ket{\varphi}.
\end{eqnarray}
Now, we return to the Eq. (\ref{decomposition}) and further require a
special case of $k_1$ and $k_2$ with $k_1=k_2=k$. Then by direct
calculations, we immediately obtain
\begin{eqnarray} \left\{
  \begin{aligned}
    a_1\varphi_1^1+a_2\varphi_2^1&=k\varphi_1\\
    a_3\varphi_3^1&=k\varphi_2
  \end{aligned}
       \right.~\text{and}~
       \left\{
       \begin{aligned}
         a_1\varphi_1^2+a_2\varphi_2^2&=k\varphi_1\\
         a_3\varphi_3^2&=k\varphi_2,
       \end{aligned}
            \right.
\end{eqnarray}
where we have used
$\ket{\varphi}=\varphi_1\ket{1}+\varphi_2\ket{2}+\varphi_3\ket{3}$. Thus, any state
$\rho$ with $\ket{\varphi_1}$ and $\ket{\varphi_2}$ and
$\widetilde{K}$ satisfy the above equations can be transformed into
the same pure state $\ket{\varphi}$ by using the stochastic incoherent
operations
\begin{eqnarray}
  \Lambda_s^1(\rho)=\frac{\widetilde{K}\rho \widetilde{K}^\dag}{\Tr(\widetilde{K}\rho \widetilde{K}^\dag)}.
\end{eqnarray}
An explicit example is
$\rho=\frac12\ket{\varphi_1}\bra{\varphi_1}+\frac12\ket{\varphi_2}\bra{\varphi_2}$, where
\begin{eqnarray}
  \ket{\varphi_1}&&=\frac1{\sqrt{2}}\left(\sin\frac{\pi}{12}\ket{1}+\cos\frac{\pi}{12}\ket{2}+\ket{3}\right)\nonumber\\
  \ket{\varphi_2}&&=\frac1{\sqrt{2}}\left(\cos\frac{\pi}{12}\ket{1}+\sin\frac{\pi}{12}\ket{2}+\ket{3}\right).
\end{eqnarray}
The corresponding incoherent Kraus operator is
\begin{eqnarray}
  \widetilde{K}=k\begin{pmatrix}
    1&1&0\\
    0& 0&1\\
    0& 0& 0
  \end{pmatrix}.
\end{eqnarray}
In other words, we can transform the $3$-dimensional mixed state
\begin{eqnarray}\label{state}
  \rho=\frac18\begin{pmatrix}
    2&1&\sqrt{6}\\
    1 & 2&\sqrt{6}\\
    \sqrt{6} & \sqrt{6}& 4
  \end{pmatrix}
\end{eqnarray}
into the pure coherent state $\ket{\varphi}$
\begin{eqnarray}
  \varphi=\frac{\widetilde{K}\rho \widetilde{K}^\dag}{\Tr(\widetilde{K}\rho \widetilde{K}^\dag)}=\frac15\begin{pmatrix}
    3&\sqrt{6}&0\\
    \sqrt{6}& 2&0\\
    0& 0& 0
  \end{pmatrix},
\end{eqnarray}
which is a pure coherent state with its coherence rank being $1$. This
incoherent Kraus operator may be came from the incoherent operation
\begin{eqnarray}
  \Lambda(\rho)=K_1\rho K_1^\dag+K_2\rho K^\dag_2,
\end{eqnarray}
where
\begin{eqnarray}
  K_1=\frac1{\sqrt{2}}\begin{pmatrix}
    1&1&0\\
    0& 0&1\\
    0& 0& 0
  \end{pmatrix},~~ K_2=\frac1{\sqrt{2}}\begin{pmatrix}
    1&-1&0\\
    0& 0&1\\
    0& 0& 0
  \end{pmatrix}.
\end{eqnarray}
Since there is no incoherent projector $\mathbb{P}$ such that, for the
state in Eq. (\ref{state}),
\begin{eqnarray}
  \frac{\mathbb{P}\rho\mathbb{P}}{\Tr(\mathbb{P}\rho\mathbb{P})}=\varphi,
\end{eqnarray}
with the coherence rank of $\ket{\varphi}$ being equal to or greater than
$1$, this completes the proof of the part (a).

(b) It is straightforward to obtain the part (b) of the theorem by
using a result in Ref. \cite{Liu1}, which says that we can transform a
mixed coherent state $\rho$ into a pure coherent state $\varphi$ with its
coherence rank $C_r(\varphi)=m\leq d-1$ by using the stochastic strictly
incoherent operations if and only if there exists an incoherent
projector $\mathbb{P}$ such that
\begin{eqnarray}
  \frac{\mathbb{P}\rho\mathbb{P}}{\Tr(\mathbb{P}\rho\mathbb{P})}=\psi, \label{theorem1}
\end{eqnarray}
with the coherence rank of $\psi$ being $n(\geq m)$. Since there is no such
incoherent projector $\mathbb{P}$ for $\rho$, thus, it is impossible to
transform $\rho$ into a pure coherent state $\varphi$ by using stochastic
strictly incoherent operations. This completes the proof of the part
(b).~~~~~~~~~~~~~~~~~~~~~~~~~~~~~~~~~~~~~~~~~~~~~~~~~~~~~~~
~~~~~~~~~~~~~~~~~~~~~~~~~~~~~~$\blacksquare$

Theorem 2 implies that the stochastic
incoherent operations are generally stronger than the stochastic strictly incoherent
operations when transform a mixed coherent state into a pure coherent
one by using them. In order to proceed further, let us recall the notion of the pure
coherent-state subspace \cite{Liu2}. If there is an incoherent
projector $\mathbb{P}$ such that $\mathbb{P}\rho\mathbb{P}= \varphi$ with the
coherence rank of $\varphi$ being $n\geq0$, then we say that $\rho$ has an
$n+1$-dimensional pure coherent-state subspace corresponding to
$\mathbb{P}$. And we say that the pure coherent-state subspace with
the projector $\mathbb{P}$ for $\rho$ is maximal if the pure
coherent-state subspace can not be expanded to a larger one with the
incoherent projector $\mathbb{P}^{\prime}$ such that
$\mathbb{P}^{\prime}\rho\mathbb{P}^{\prime}= \varphi^{\prime}$,
$\varphi^{\prime}\neq\varphi$, and
$\mathbb{P}\varphi^{\prime}\mathbb{P}= \varphi$. From \emph{Theorem} 2, we
obtain that the stochastic incoherent operations can increase the the dimension of the maximal pure-coherent subspace.

Next, with the above notions, let us consider the operational difference
between incoherent operations and strictly incoherent operations.  From
 corollary 1, we know that there is no operational gap between them in
 the $2$-dimensional case. Thus, let us consider the $3$-dimensional
 case. This leads the following theorem.

\emph{Theorem} 3.-- Let $\rho$ be any $3$-dimensional mixed state with
its the dimension of the maximal pure-coherent subspace being $1$. Then incoherent
 operations cannot increase the dimension of the maximal pure-coherent subspace of $\rho$.

\emph{Proof.}--To prove the theorem, let us consider the transformation
\begin{eqnarray}\label{3dimensional}
\Lambda(\rho)=\varphi,
\end{eqnarray}
where $\Lambda(\cdot)=\sum_nK_n(\cdot)K_n^\dag$ is some incoherent
operation and $\ket{\varphi}=(m,n,0)^t$ is a $3$-dimensional pure coherent
state with $\abs{m}^2+\abs{n}^2=1$. Next, we will divide our discussion
into four cases.

Case a) There are three linearly independent incoherent kraus operators as
\begin{eqnarray}\label{type}
K=\begin{pmatrix}
    a&b&c\\
    0& 0&0\\
    0& 0& 0
  \end{pmatrix},
\end{eqnarray}
with $a, b, c\neq 0$, up to a permutation  in $\{K_n\}$. Without loss of generality,
we assume them as
\begin{eqnarray}\nonumber
  K_1=\begin{pmatrix}
    a_1&a_2&a_3\\
    0& 0&0\\
    0& 0& 0
  \end{pmatrix},~~ K_2=\begin{pmatrix}
    b_1&b_2&b_3\\
    0& 0&0\\
    0& 0& 0
  \end{pmatrix},~~ K_3=\begin{pmatrix}
    c_1&c_2&c_3\\
    0& 0&0\\
    0& 0& 0
  \end{pmatrix}.
\end{eqnarray}
The relation in Eq. (\ref{3dimensional}) implies that there are at least two $3$-dimensional
states $\ket{\psi_1}=(x_1,x_2,x_3)^t$ and $\ket{\psi_2}=(y_1,y_2,y_3)^t$ such that
\begin{eqnarray}
K_n\ket{\psi_i}=\textbf{0},
\end{eqnarray}
for $n=1,2,3$ and $i=1,2$, where $\textbf{0}$ is the null vector.
This means that there are
\begin{eqnarray}
a_1x_1+a_2x_2+a_3x_3&&=0,\nonumber\\
b_1x_1+b_2x_2+b_3x_3&&=0,\nonumber\\
c_1x_1+c_2x_2+c_3x_3&&=0.\label{equation3}
\end{eqnarray}
Since $K_1, K_2,K_3$ are linearly independent, then the rank of the matrix
\begin{eqnarray}
A=\begin{pmatrix}
    a_1&a_2&a_3\\
    b_1& b_2&b_3\\
    c_1&c_2& c_3,
  \end{pmatrix}
\end{eqnarray}
is $3$. We immediately obtain the uniqueness of   $\ket{\psi}=(x_1,x_2,x_3)^t$.

Case b) There are two linearly independent incoherent kraus operators as Eq. (\ref{type})
in $\{K_n\}$. Without loss of generality,
we assume them as (up to a permutation)
\begin{eqnarray}\nonumber
  K_1=\begin{pmatrix}
    a_1&a_2&a_3\\
    0& 0&0\\
    0& 0& 0
  \end{pmatrix},~~ K_2=\begin{pmatrix}
    b_1&b_2&b_3\\
    0& 0&0\\
    0& 0& 0
  \end{pmatrix},
\end{eqnarray}
with $a_i,b_i\neq0$ for all $i$ and $j$.
The relation in Eq. (\ref{3dimensional}) implies that
\begin{eqnarray}\label{equation4}
a_1x_1+a_2x_2+a_3x_3&&=0,\nonumber\\
b_1x_1+b_2x_2+b_3x_3&&=0.
\end{eqnarray}
From Eq. (\ref{equation4}) and $\abs{x_1}^2+\abs{x_2}^2+\abs{x_3}^2=1$,
by direct calculations, we immediately obtain the uniqueness of   $\ket{\psi}=(x_1,x_2,x_3)^t$.

Case c) There is only one incoherent Kraus operator as Eq. (\ref{type}) in $\{K_n\}$.
Without loss of generality, we assume it as
\begin{eqnarray}\nonumber
  K_1=\begin{pmatrix}
    a_1&a_2&a_3\\
    0& 0&0\\
    0& 0& 0
  \end{pmatrix},
\end{eqnarray}
with $a_i\neq0$ for all $i$.
Since $\Lambda(\cdot)$ is a trace preserving map, then there must be three incoherent
 Kraus operators in  $\{K_n\}$ with the following forms:
\begin{eqnarray}\nonumber
  K_1^\prime=\begin{pmatrix}
    a_1^\prime&a_2^\prime&0\\
    0& 0&a_3^\prime\\
    0& 0& 0
  \end{pmatrix},~~ K_2^\prime=\begin{pmatrix}
    b_1^\prime&0&0\\
    0& b_2^\prime&b_3^\prime\\
    0& 0& 0
  \end{pmatrix},~~ K_3^\prime=\begin{pmatrix}
    c_1^\prime&0&c_3^\prime\\
    0& c_2^\prime&0\\
    0& 0& 0
  \end{pmatrix}.
\end{eqnarray}
Then, $(a_1,a_2)^t$ and $(a_1^\prime,a_2^\prime)$,  $(a_2,a_3)^t$ and
 $(b_2^\prime,b_3^\prime)$, and $(a_1,a_3)^t$ and $(c_1^\prime,c_3^\prime)$
 are all linearly independent.
The relation in Eq. (\ref{3dimensional}) implies that
\begin{eqnarray}\label{equation6}
a_1x_1+a_2x_2+a_3x_3&&=0.
\end{eqnarray}
and
\begin{eqnarray}\label{equation7}
(a_1^\prime x_1+a_2^\prime x_2)n&&=a_3^\prime x_3 m,\nonumber\\
b_1^\prime x_1 n&&=(b_2^\prime x_2+b_3^\prime x_3)m,\nonumber\\
(c_1^\prime x_1+c_3^\prime x_3)n&&=c_2^\prime x_2 m.
\end{eqnarray}
From Eqs. (\ref{equation6}) and (\ref{equation7}), by direct calculations, we
immediately obtain the uniqueness of   $\ket{\psi}=(x_1,x_2,x_3)^t$.

Case d)  There is no incoherent Kraus operator  as Eq. (\ref{type}) in $\{K_n\}$.
Since $\Lambda(\cdot)$ is a trace preserving map and strictly incoherent operations
cannot transform a $3$-dimensional mixed states into a pure coherent states
\cite{Liu1}, then there is at least a pair of incoherent Kraus operators both
from one of the following three types:
\begin{eqnarray}\nonumber
  K_1^\prime=\begin{pmatrix}
    a_1^\prime&a_2^\prime&0\\
    0& 0&a_3^\prime\\
    0& 0& 0
  \end{pmatrix},~~ K_2^\prime=\begin{pmatrix}
    b_1^\prime&0&0\\
    0& b_2^\prime&b_3^\prime\\
    0& 0& 0
  \end{pmatrix},~~ K_3^\prime=\begin{pmatrix}
    c_1^\prime&0&c_3^\prime\\
    0& c_2^\prime&0\\
    0& 0& 0
  \end{pmatrix}.
\end{eqnarray}
Without loss of generality, let them all have the form as $K_1^\prime$ with
\begin{eqnarray}\nonumber
  K_1=\begin{pmatrix}
    a_1&a_2&0\\
    0& 0&a_3\\
    0& 0& 0
  \end{pmatrix},~~ K_2=\begin{pmatrix}
    b_1&b_2&0\\
    0& 0&b_3\\
    0& 0& 0
  \end{pmatrix}.
\end{eqnarray}
Then the relations in Eq. (\ref{3dimensional}) and $\abs{x_1}^2+\abs{x_2}^2+\abs{x_3}^2=1$ imply that
\begin{eqnarray}\label{equation8}
(a_1x_1+a_2x_2)n&&=a_3x_3m,\nonumber\\
(b_1x_1+b_2x_2)n&&=b_3x_3m.
\end{eqnarray}
From Eq. (\ref{equation8}), by direct calculations, we immediately obtain the uniqueness of   $\ket{\psi}=(x_1,x_2,x_3)^t$.

Thus, the above four cases imply that incoherent operations cannot increase the dimension of the maximal pure-coherent subspace of a $3$-dmensional $\rho$.
This completes the proof of the theorem.
~~~~~~~~~~~~~~~~~~~~~~~~~~~~~~~~~~~~~~~~~~~~~~~~~~~~~~~~~~~~~~~~~~~~~~~~~~~~~~~~~~~~$\blacksquare$

Theorem 3 tells us that incoherent operations cannot increase the dimension of the maximal pure-coherent subspace of a $3$-dimensional $\rho$.
Let us consider the $4$-dimensional state space
and show that incoherent operations can increase the dimension of the maximal pure-coherent subspace of a state.

\emph{Theorem} 4.--Let $\rho$ be a $4$-dimensional mixed state with its dimension of the maximal
pure-coherent subspace being $1$. There is an incoherent operation such that the transformation from
$\rho$ to $\varphi$, with its coherence rank $\geq1$, can be achieved with certainty.

\emph{Proof.}--To this end, let us consider the incoherent operation $\Lambda(\cdot)$ which has the form
\begin{eqnarray}\label{map}
\Lambda(\cdot)=K_1(\cdot)K^\dag_1+K_2(\cdot)K^\dag_2
\end{eqnarray}
with
\begin{eqnarray}\label{K_1}
  K_1=\begin{pmatrix}
    a&b&0&0\\
    0&0&c&d\\
    0& 0&0&0\\
    0& 0& 0&0
  \end{pmatrix},~~
 K_2=\begin{pmatrix}
    -b&a&0&0\\
    0&0&-d&c\\
    0& 0&0&0\\
    0& 0& 0&0
  \end{pmatrix}.
\end{eqnarray}
The relation $K_1^\dag K_1+K_2^\dag K_2=I$ implies that $\abs{a}^2+\abs{b}^2=1$ and $\abs{c}^2+\abs{d}^2=1$.
Let $\rho=p_1\ket{\varphi_1}\bra{\varphi_1}+p_2\ket{\varphi_2}\bra{\varphi_2}$. Then, the deterministic transformation
\begin{eqnarray}
\Lambda(\rho)=\varphi
\end{eqnarray}means that, for $i=1,2$, there are
\begin{eqnarray}
K_1\ket{\varphi_i}&&=k_i^1\ket{\varphi}~\text{or}~\textbf{0},\nonumber\\
K_2\ket{\varphi_i}&&=k_i^2\ket{\varphi}~\text{or}~\textbf{0}.
\end{eqnarray}
Specifically, we assume that there are
\begin{eqnarray}\label{relation}
K_1\ket{\varphi_1}&&=k_1^1\ket{\varphi}~\text{and}~K_1\ket{\varphi_2}=\textbf{0},\nonumber\\
K_2\ket{\varphi_1}&&=\textbf{0}~\text{and}~K_2\ket{\varphi_2}=k_2^2\ket{\varphi}.
\end{eqnarray}
Equipped with these tools, we may construct an explicit example as follows.
Let $\ket{\varphi}=\frac1{\sqrt{2}}(\ket{0}+\ket{1})$, $a=\frac45$, $b=\frac35$, $c=\frac1{\sqrt{5}}$, and $d=\frac2{\sqrt{5}}$.
Then, from Eq. (\ref{relation}), by direct calculations, $\ket{\varphi_1}$ and $\ket{\varphi_2}$ can be chosen as
\begin{eqnarray}
\ket{\varphi_1}&&=\frac1{5\sqrt{2}}(4,3,\sqrt{5},2\sqrt{5})^t\\
\ket{\varphi_2}&&=\frac1{5\sqrt{2}}(-3,4,-2\sqrt{5},\sqrt{5})^t,
\end{eqnarray}
where the superscript $t$ means transpose.
Without loss of generality, we further assume that $p_1=p_2=\frac12$.

Thus, from the above discussion, the state we chosen is
\begin{eqnarray}\label{state}
  \rho=\begin{pmatrix}
    \frac14&0&\frac1{2\sqrt{5}}&\frac1{4\sqrt{5}}\\
    0&\frac14&-\frac1{4\sqrt{5}}&\frac1{2\sqrt{5}}\\
   \frac1{2\sqrt{5}}&-\frac1{4\sqrt{5}}&\frac14&0\\
    \frac1{4\sqrt{5}}&\frac1{2\sqrt{5}}& 0&\frac14
  \end{pmatrix},
\end{eqnarray}
and the incoherent operation we chosen is
\begin{eqnarray}
  \Lambda(\rho)=K_1\rho K_1^\dag+K_2\rho K_2^\dag,
\end{eqnarray}
with its incoherent Kraus operators being
\begin{eqnarray}\label{K_1}
  K_1=\begin{pmatrix}
    \frac45&\frac35&0&0\\
    0&0&\frac1{\sqrt{5}}&\frac2{\sqrt{5}}\\
    0& 0&0&0\\
    0& 0& 0&0
  \end{pmatrix},~~
 K_2=\begin{pmatrix}
    -\frac35&\frac45&0&0\\
    0&0&-\frac2{\sqrt{5}}&\frac1{\sqrt{5}}\\
    0& 0&0&0\\
    0& 0& 0&0
  \end{pmatrix}.
\end{eqnarray}
By direct calculations, from Eqs. (\ref{state}) and (\ref{K_1}), we immediately obtain that
\begin{eqnarray}
  \Lambda(\rho)=\frac12\begin{pmatrix}
    1&1&0&0\\
    1&1&0&0\\
   0&0&0&0\\
    0&0& 0&0
  \end{pmatrix}.
\end{eqnarray}
It is straightforward to examine that the dimension of the maximal
pure coherent-state subspace of $\rho$ is $1$, the dimension of the maximal pure
coherent-state subspace of $\Lambda_s(\rho)$ is $2$, and the corresponding
incoherent projector is
\begin{eqnarray}
  \mathbb{P^\prime}=\ket{1}\bra{1}+\ket{2}\bra{2}.
\end{eqnarray}
Thus, we have found a mixed
state $\rho$ with its dimension of the maximal pure coherent-state
subspace of it being $1$ while, for some incoherent
operations $\Lambda(\cdot)$, the dimension of the maximal pure coherent-state subspace of
$\Lambda(\rho)$ being $n(>1)$.
This completes the proof of the theorem.
~~~~~~~~~~~~~~~~~~~~~~~~~~~~~~~~~~~~~~$\blacksquare$

At last, we would like to present two applications of our results.

The first application is related
 to the the operational gap between incoherent operations and strictly
incoherent operations. As we know,  although the inclusion relations
between them is known as $\text{PIO}\subset \text{SIO}\subset \text{IO} \subset \text{MIO}$,
the operational gap in terms of state transformation between them has attracted much attentions
\cite{Chitambar,Streltsov}. Specifically, the operational gap of maximally incoherent operations
and the incoherent operations under stochastic state transformation is presented in Refs.
\cite{Chitambar1,Fang}, the operational gap between strictly incoherent operations and
physically incoherent operations under stochastic state transformation is presented in
Ref. \cite{Chitambar2}. However, whether there is an operational gap between incoherent
operations and strictly incoherent operations under deterministic state transformation is unclear
and is an open question \cite{Chitambar}. Here, from our Theorem 4, we see that there is indeed an
operational gap between the incoherent operations and the strictly incoherent operations
under deterministic state transformation.

 Another application is related to the difference between
coherence monotones under IO and that of SIO. Theorem 4
leads directly to the existence of coherence monotones under
strictly incoherent operations that increase under incoherent operations. As we known, any
deterministic transformation not achievable by strictly incoherent operations necessarily
implies the increase in some coherence monotone. Thus, we arrive at the following corollary.

\emph{Corollary 2.}--There exist coherence monotones under strictly incoherent operations that can
increase under incoherent operations.

The results of Ref. \cite{Liu2} show that the dimension of the maximal pure coherent-state
subspace cannot increase under strictly incoherent operations. However, from Theorem 4, we
immediately obtain that  the dimension of the maximal pure coherent-state
subspace of a state can increase under incoherent operations.

\section{Conclusions}

In summary, we have investigated the transformation from a mixed
coherent state into a pure one by using both the incoherent operations
and the stochastic incoherent operations. Our results show that contrary
to the strictly incoherent operations and the stochastic strictly incoherent
operations, both the incoherent operations and the stochastic incoheren
operations can increase the dimension of the maximal pure-coherent subspace of a state. This
means that the
incoherent operations are generally stronger than the strictly incoherent
operations when transform a mixed coherent state into a pure coherent
one by using them. Our findings confirmed that there is an operational gap
between incoherent operations and strictly incoherent operations
under state transformations, which provide an answer to the open question in
Ref. \cite{Chitambar1}. As an application, we have further shown that  there exist
coherence monotones under strictly incoherent operations that can increase
under incoherent operations. The subtle differences between the incoherent operations
and the strictly incoherent operations will increase our
understandings on quantum coherence transformations under free operations
in the resource theory of coherence.

\section*{Acknowlegements}
The authors acknowledge financial supported from NSF of China (Grant
No.11775300 and No.12075310), the National Key Research and Development Program of
China (2016YFA0300603),  the Strategic Priority Research Program of
Chinese Academy of Sciences No. XDB28000000, and the China
Postdoctoral Science Foundation Grant No. 2019M660841.

\end{document}